\documentclass[12pt]{iopart}

\usepackage{iopams}  

\usepackage{cite}
\usepackage{graphicx}%
\usepackage{hyperref}
\usepackage[dvipsnames]{xcolor}
\usepackage{ulem}

\begin{document}

\title{
Itinerant magnetism of chromium under pressure: a DFT+DMFT study
}

\author{A S Belozerov$^{1}$, A A Katanin$^{2,1}$, V I Anisimov$^{1,3}$}

\address{$^1$ M. N. Miheev Institute of Metal Physics, Russian Academy of Sciences, 620108 Yekaterinburg, Russia}
\address{$^2$ Center for Photonics and 2D Materials, Moscow Institute of Physics and Technology, 141701 Dolgoprudny, Russia}
\address{$^3$ Ural Federal University, 620002 Yekaterinburg, Russia}

\begin{abstract}
We consider electronic and magnetic properties of chromium, a well-known itinerant antiferromagnet, by a combination of density functional theory (DFT) and dynamical mean-field theory (DMFT). We find that electronic correlation effects in chromium, in contrast to its neighbours in the periodic table, are weak, leading to the quasiparticle mass enhancement factor ${m^*/m \approx 1.2}$. Our results for local spin-spin correlation functions and distribution of weigths of atomic configurations indicate that the local magnetic moments are not formed. Similarly to previous results of DFT at ambient pressure, the non-uniform magnetic susceptibility as a function of momentum possesses close to the wave vector ${{\mathbf Q}_{\rm H}=(0,0,2\pi/a)}$ ($a$ is the lattice constant) sharp maxima, corresponding to Kohn anomalies. We find that these maxima are preserved by the interaction and are not destroyed by pressure. Our calculations qualitatively capture a decrease of the N\'eel temperature with pressure and a breakdown of itinerant antiferomagnetism at pressure of $\sim$9~GPa in agreement with experimental data, although the N\'eel temperature is significantly overestimated because of the mean-field nature of DMFT.

\end{abstract}
\maketitle

\section{Introduction}

Chromium represents a well-known magnet possessing spin-density-wave order and having substantial N\'eel temperature $T_{\rm N}\approx 300$~K at ambient pressure, see review \cite{Fawcett} and references therein. The magnetic order in chromium was shown to be incommensurate with the wave vector ${\bf Q}=(2\pi/a)(0,0,1-\delta)$ at not too low temperatures ($a$ is the lattice constant) with weakly temperature dependent incommensurability $\delta\sim 0.05$ \cite{Neutron}. Already in early study of Ref. \cite{Bridgman} it was noted that the anomaly of resistivity, which occurs at the N\'eel temperature, is shifted to lower temperatures with applied pressure.  Recent studies at high pressures \cite{Pressure1,Pressure2} have shown that the antiferromagnetism is (almost) fully suppressed by pressure, and at a critical pressure ${p_c\approx 9.5}$~GPa the quantum phase transition to paramagnetic state occurs.

Soon after the discovery of antiferromagnetism of chromium it was proposed that it occurs due to nesting of the Fermi surface \cite{Overhauser,Lomer}. Despite the simplicity of the corresponding mean-field theory, the non-local correlations yield non-trivial physical effects \cite{Dzyaloshinskii,Dzyaloshinskii1,fRG}.
It was however emphasized \cite{Roth,Rice} that imperfectness of the nesting is an important factor, which changes essentially physical picture of the origin of spin density wave. Namely, a cusp-type maximum of the non-uniform susceptibility at some wave vector ${\bf Q}$ may arise due to the so-called Kohn points of the Fermi surface separated by the wave vector ${\bf Q}$. The Kohn points are the points, having opposite Fermi velocities, and they yield local maximum of non-uniform susceptibility at the momentum $\mathbf Q$ if the signs of the effective electronic masses at these points are opposite in two perpendicular directions \cite{Rice,Roth}. Due to electronic interaction, this maximum, having non-analytic momentum dependence, may yield spin density wave with the wave vector~$\bf Q$. It was argued recently that Kohn points may form lines, located on the Fermi surfaces\cite{OurKohn,OurCr}. These lines are preserved by the interaction effects \cite{OurKohn}, and seem to be present in chromium \cite{OurCr,OurCr1}.

The electronic structure of Cr has been extensively studied within density functional theory (DFT) \cite{Rath,Fry,Fry1,Chen1988,Singh1992,Guo2000,
Hafner2002
}. In particular, the shape of the Fermi surface sheets was studied and their approximately nested parts were identified by \textit{ab initio} methods \cite{Rath,Fry,Fry1,Hafner2002}. Apart from that, the momentum dependence of particle-hole bubble was investigated \cite{Susc0,Susc1} and  
magnetic ground state was described~\cite{Staunton1999,Cottenier2002, Soulairol2010}.
However, the DFT alone does not give a possibility to extract magnetic transition temperatures. It also has difficulties in description of paramagnetic state and electron correlation effects arising in partially-filled $d$ bands.
An accurate treatment of many-body effects, including correlations at finite temperature,
can be performed, e.g., by a combination of DFT and dynamical mean-field theory (DMFT)~\cite{dmft,dmft1}.
This combination, called DFT+DMFT~\cite{dftdmft}, was previously applied to describe magnetism of iron 
\cite{Fe,alpha_iron2010,Fe_leonov, OurAlphaIgoshevKatanin,OurAlphaBelozerovKatanin,Hausoel}, nickel \cite{Fe,Hausoel}, and
recently to ZrZn$_2$, a prototypical weak ferromagnet~\cite{Skornyakov_ZrZn2}. 
These studies showed that the effect of electronic interaction is essential for 
the formation of local magnetic moments and their screening.

In this paper, we study the electronic and magnetic properties of paramagnetic Cr at pressures up to 10 GPa within the DFT+DMFT approach. In contrast to the weak ferromagnet ZrZn$_2$ we find that magnetism in chromium is purely itinerant and no local magnetic moments are formed. Using the supercell approach, we extract the N\'eel temperature, which decreases with 
pressure, but found to be overestimated because of mean-field nature of DMFT approach. We find a breakdown of antiferromagnetism at 9~GPa in agreement with experimental data. However, the obtained quantum phase transition is of the first kind, which, as we argue, reflects deficiency of supercell approach.

\section{Computational details}
\label{sec:computational_details}

We have performed DFT calculations using the pseudopotential plane-wave method implemented in the Quantum-ESPRESSO package~\cite{ESPRESSO}.
The Vanderbilt ultrasoft pseudopotential 
with the Perdew-Burke-Ernzerhof form of generalized gradient approximation was used.
The convergence threshold for total energy was set to $10^{-6}$~Ry.
The integration in the reciprocal space was performed using 20$\times$20$\times$20 $\textbf{k}$-point mesh in all calculations except those of momentum-dependent susceptibility, where 40$\times$40$\times$40 mesh was employed.
The calculations were carried out with the experimental lattice constants taken at the corresponding pressures~\cite{Pressure1}.
We also have computed the pressure dependence of the lattice constant within DFT by fitting to the third-order Birch-Murnaghan equation of state. The obtained lattice constant is found to be less than the experimental one by 0.035~$\AA$ at ambient pressure and 0.018~$\AA$ at the pressure of 10~GPa. We have checked that this difference does not qualitatively affect our results.
For DMFT calculations we have constructed a basis of maximally-localised Wannier functions (MLWFs)~\cite{MLWF} 
by means of Wannier90 code~\cite{wannier90}.
To take into account the hybridization of $3d$ states with $4s$ and $4p$ ones, we include all of them in our Wannier function basis.

We parametrize the Coulomb interaction for $d$ shell via Slater integrals $F^0$, $F^2$, and $F^4$ linked to the Hubbard parameter ${U\equiv F^0}$ and Hund's rule coupling ${J_{\rm H}\equiv (F^2+F^4)/14}$ 
(details can be found in reference~\cite{u_and_j}).
In our calculations we adopt 
${J_{\rm H}=0.9}$~eV, which is commonly used for $3d$~metals~\cite{Anisimov_book,Fe,Fe_leonov,alpha_iron2010,OurAlphaIgoshevKatanin,OurAlphaBelozerovKatanin,BelozerovSU2}.
We also adopt ${U=2.5}$~eV, which is close to ${U\approx U_{\rm eff} + J_{\rm H} = 2.9 \pm 0.4}$~eV extracted from the resonant photoemission spectra~\cite{Kaurila1997}
and ${U\sim 2.8-3}$~eV obtained by the constrained random-phase approximation in the MLWFs basis~\cite{footnote_U, Miyake2008}.

Our DMFT calculations have been performed using the AMULET code~\cite{amulet}.
To account for the electronic interactions already described by DFT, we use the around mean-field form of double-counting correction, evaluated from the self-consistently determined local occupations.
We also verified that the fully localized form of double-counting correction leads to similar results with a slightly less ($\sim$0.04) filling of \textit{d} states.
To compute the density of states, we perform the analytical continuation of self-energy to real-energy range by using the Pad\'e approximants~\cite{Pade}.

The impurity problem has been solved by the hybridization expansion continuous-time quantum Monte Carlo method~\cite{CT-QMC} with the density-density form of Coulomb interaction.
This form of Coulomb interaction
is employed in most part of material-specific DMFT calculations and corresponds to the Ising ($Z_2$) symmetry of Hund's exchange. Although this approximation affects the electronic properties near the Mott transition~\cite{Antipov2012} and results in overestimation of the Curie temperature in strong magnets\cite{Antipov2012,Hausoel,BelozerovSU2}, it drastically reduces the computational costs making such calculations feasible. Moreover, it yields qualitatively- and semiquantitatively correct results for strong magnets, and expected to be applicable to the weak magnets as well.
%
%

\section{Results and discussion}
\subsection{Electronic and local magnetic properties\label{Sect31}}

\begin{figure}[t]
\centering
\includegraphics[clip=true,width=0.53\textwidth]{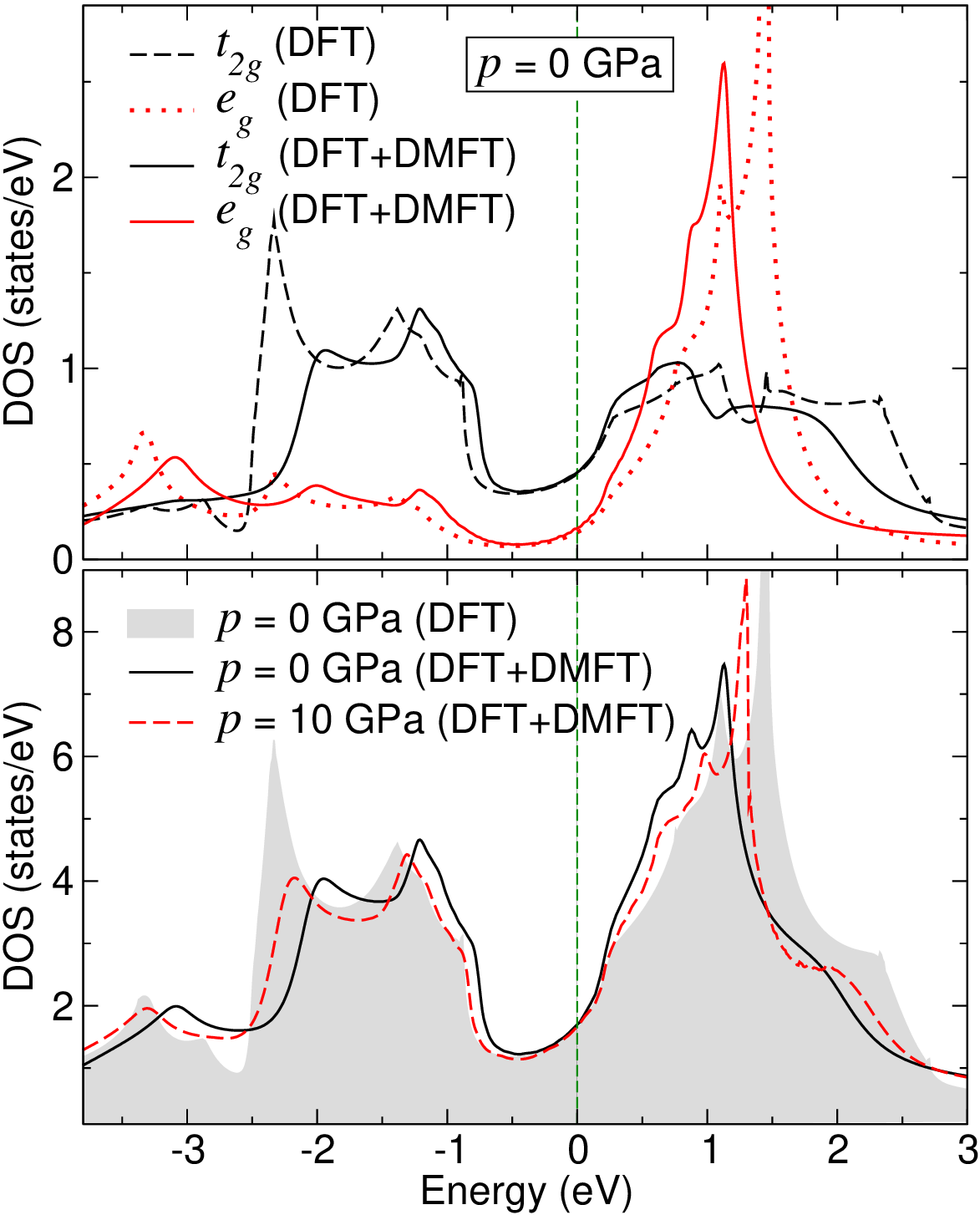}
\caption{\label{dos}
Orbital projected density of $3d$ states (DOS) at ambient pressure (labeled as $p = 0$~GPa, top panel) and total DOS (bottom panel) at ambient pressure and ${p=10}$~GPa as obtained within non-spin-polarised DFT method and DFT+DMFT method at temperature ${T = 193}$~K. The Fermi level is at zero energy.}
\end{figure}

\begin{figure}[t]
\centering
\includegraphics[clip=true,width=0.52\textwidth]{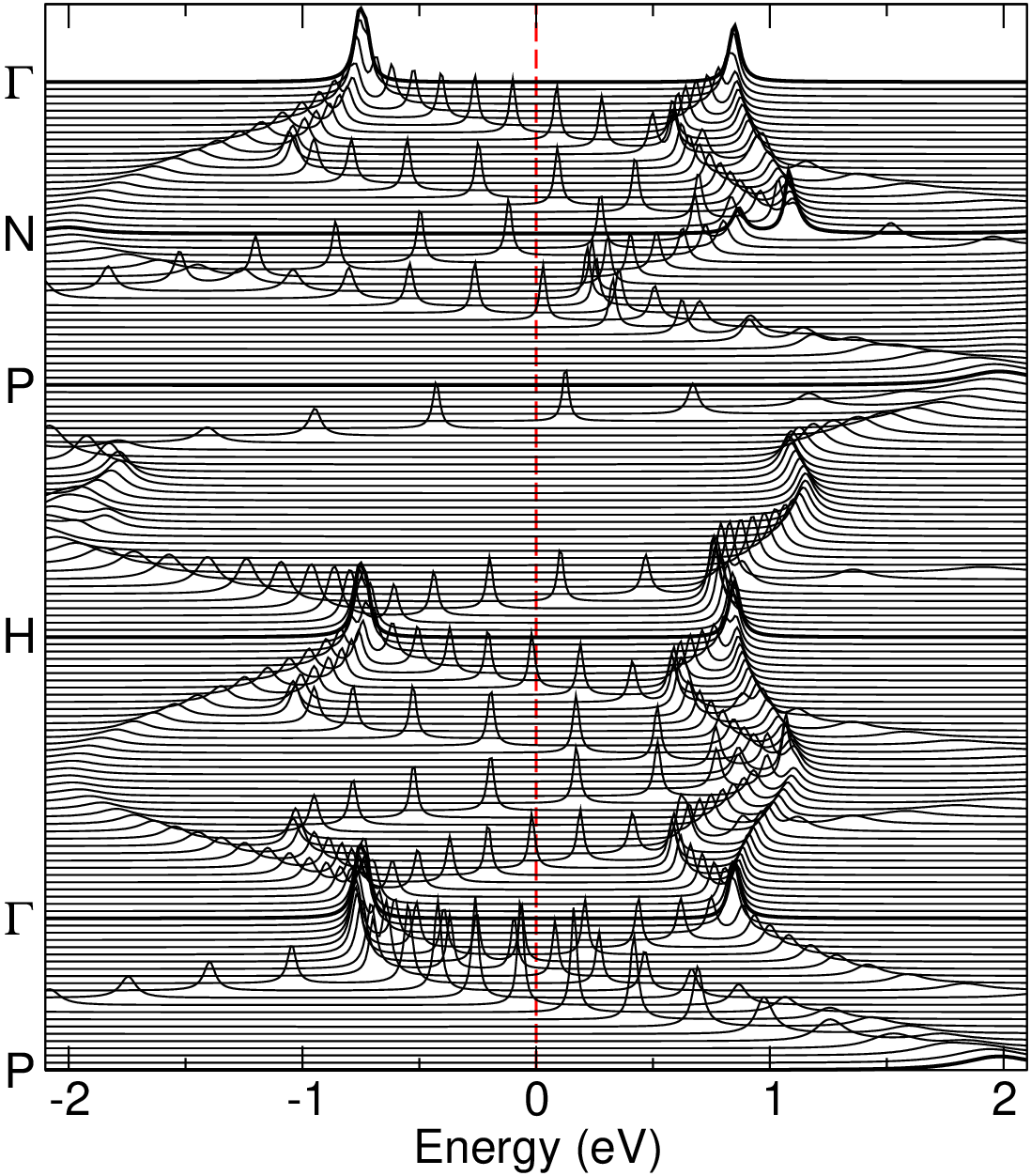}
\caption{\label{fig:akw}
Momentum resolved spectral function obtained within DFT+DMFT method at ambient pressure and temperature ${T = 193}$~K. The Fermi level is at zero energy.}
\end{figure}

Our DFT+DMFT calculations yield the $d$-states filling of 4.02, which is almost independent of pressure up to 10~GPa.
At the same time, orbitals of $t_{2g}$ and $e_g$ symmetry have different fillings of 0.91 and 0.64, respectively.
The former is not far from half-filling, that may lead to significant correlation effects~\cite{spin_freezing}.
However, the density of states (DOS), shown in figure~\ref{dos}, is low at the Fermi level for states of both symmetry.
Moreover, the peak in density of $e_g$ states is relatively far from the Fermi level (comparing, e.g., to $\alpha$-iron, where it plays an important role in formation of well-localized magnetic moments~\cite{alpha_iron2010,OurAlphaIgoshevKatanin,OurAlphaBelozerovKatanin}).
As seen in figure~\ref{dos}, dynamical correlation effects considered in DMFT lead to a slight renormalization of states near the Fermi level, decreasing the distance to the $e_g$ peak to 1.1~eV compared to 1.4~eV in DFT at ambient pressure.
As seen in the bottom panel of figure 1, the shape of the DOS obtained within DFT+DMFT is weakly affected by pressure of 10~GPa, albeit a slight shift of the position of the peaks and broadening of the bandwidth is observed.

\begin{figure}[t]
\hspace{7.85em}
\includegraphics[clip=true,width=0.55\textwidth]{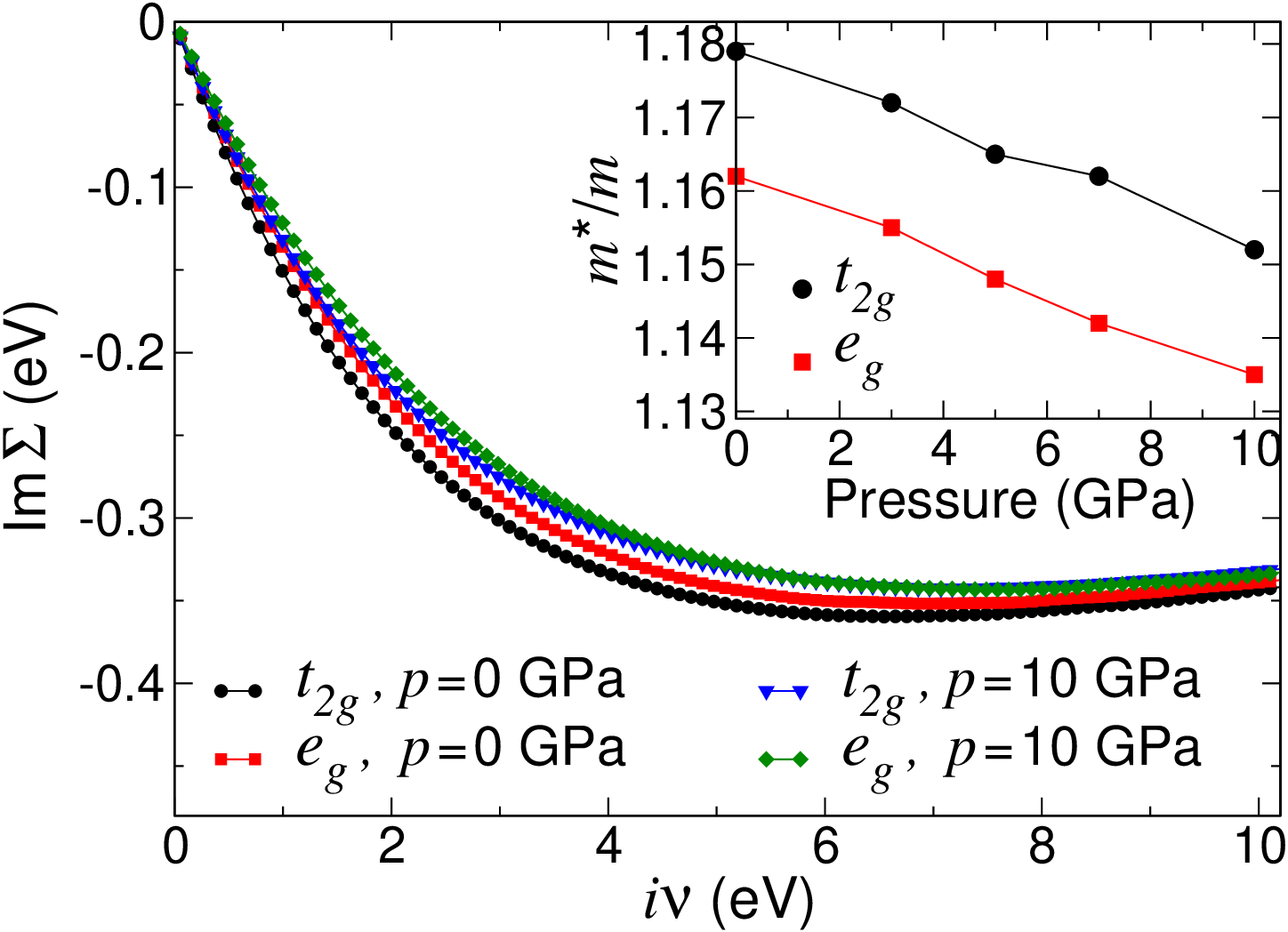}
\caption{ \label{sigma}
Imaginary part of electronic self-energy $\Sigma$ as a function of imaginary frequency $i\nu$ (main panel) obtained by DFT+DMFT at pressures ${p=0, 10}$~GPa and temperature ${T = 193}$~K. Inset: quasiparticle mass enhancement ${m^*/m}$ as a function of pressure.}
\end{figure}

In figure~\ref{fig:akw} we present momentum resolved spectral functions $A({\bf k},\nu)=-(1/\pi)\sum_m {\rm Im} G^{mm}_{\bf k}(\nu)$ where 
${\bf k}$ is the momentum,
$G_{\bf k}(\nu)$ is the one-particle Green's function, which is a matrix in the Wannier function space. The Green's function was obtained using the Wannier-projected Hamiltonian and self-energy continued to the real frequency axis using P\'ade approximants. One can see that near the Fermi level the spectral functions have a form of relatively narrow peaks, corresponding to well defined quasiparticles.  

In figure~\ref{sigma} we present the imaginary part of electronic self-energy $\Sigma$ as a function of imaginary frequency $i\nu$ at ambient pressure and ${p=10}$~GPa.
The obtained frequency dependencies of the self-energy have a Fermi-liquid-like form with a small quasiparticle damping (i.e., inverse quasiparticle lifetime), and have a similar shape and magnitude at temperatures 193 and 1160~K.
To estimate the strength of electronic correlations, we
compute the quasiparticle mass enhancement $m^*/m\,{=}\,1 - \left[d\,{\rm Im} \Sigma (i\nu)/ d\nu \right]_{\nu\to 0}$ using Pad\'e approximants for self-energy.
At ambient pressure and temperature of 193~K we obtain ${m^*/m}$ of 1.18 and 1.16 for $t_{2g}$ and $e_g$ states, respectively. 
Upon compression of the lattice, the calculated ratio ${m^*/m}$ decreases monotonically as shown in the inset of figure~\ref{sigma}, that can be explained by an increase of the bandwidth.
We have also checked that the correlation effects weakly depend on temperature. In particular, the difference of averaged ${m^*/m}$ at these temperatures is only about 0.01.
The obtained values of ${m^*/m}$ indicate that correlation effects in chromium are rather weak.

\begin{figure}[t]
\hspace{8.5em}
\includegraphics[clip=true,width=0.59\textwidth]{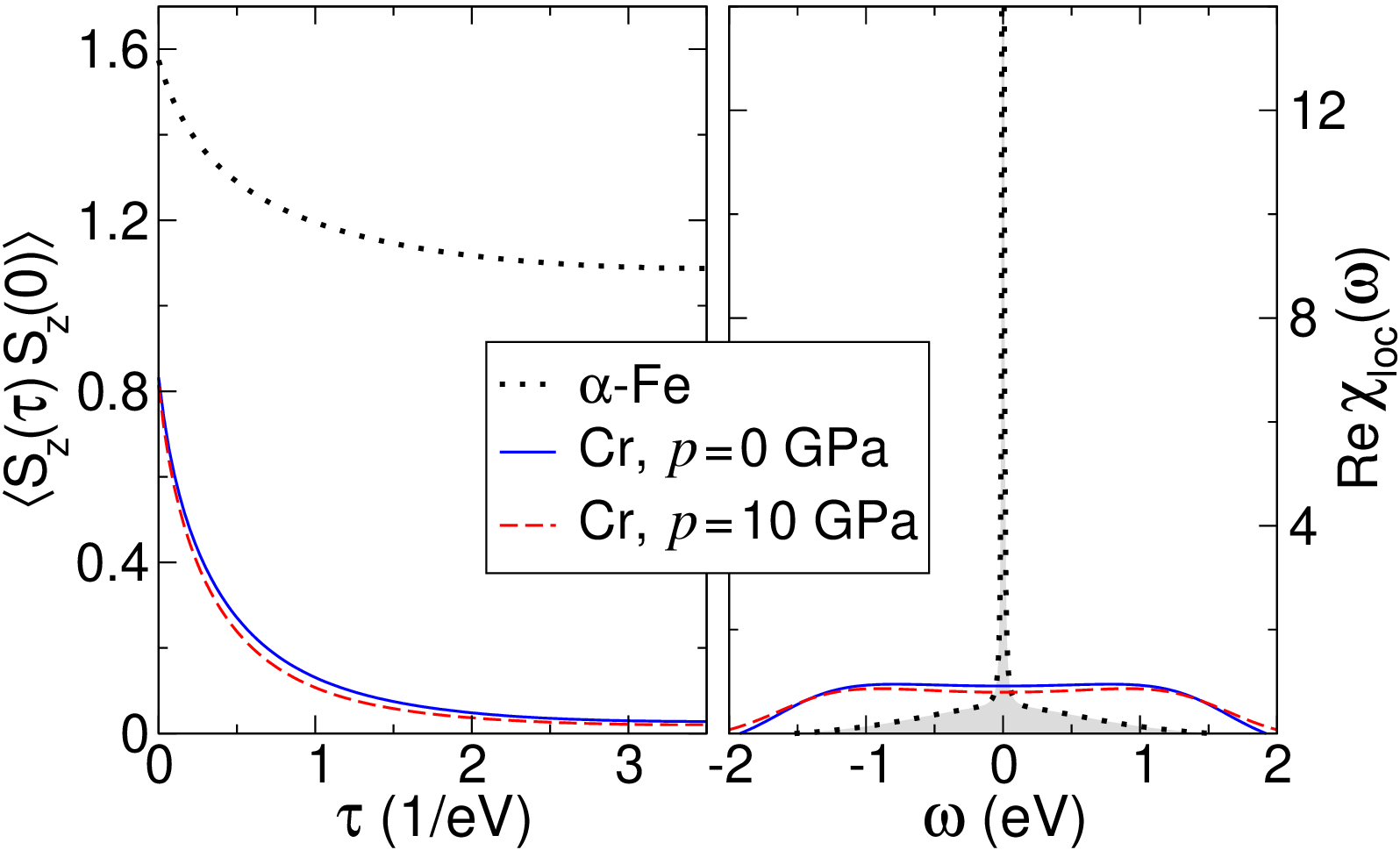}
\caption{ \label{fig:correlators}
Local spin-spin correlation functions in the imaginary-time (left panel) and real-frequency (right panel) domains calculated by DFT+DMFT method at inverse temperature ${\beta=7}$~eV$^{-1}$. The obtained correlation functions for chromium at pressures ${p = 0}$ and ${p = 10}$~GPa are compared with that for $\alpha$-Fe at ambient pressure~\cite{footnote1}.}
\end{figure}

This is also confirmed by the study of local dynamic susceptibility,
expressed as a correlation function of local spin operators
$\chi_{\rm loc}(\tau) = \langle S_z(\tau) S_z(0) \rangle$, where $S_z$ is the $z$-component of the
local spin operator and $\tau$ is the imaginary time.
In figure~\ref{fig:correlators} we show the dependence $\chi_{\rm loc}(\tau)$ at inverse temperature ${\beta=1/T=7}$~eV$^{-1}$, together with the real part of $\chi_{\rm loc}(\omega)$, obtained by Fourier transform and analytical continuation to real frequency~$\omega$. One can see that the local susceptibility shows strong imaginary time dependence, which drops almost to zero at ${\tau \sim \beta/2}$ and the corresponding plateau in real frequency dependence, which shows absence of local magnetic moments (cf. the corresponding dependencies for such strong magnet, as iron). Accordingly, the local static susceptibility $\chi_{\rm loc}^0 = \int_0^\beta {\chi_{\rm loc}(\tau) d\tau}$ (not shown) is only weakly temperature dependent. Therefore, one can characterize chromium as a weak itinerant magnet. 

\begin{figure}[t]
\centering
\includegraphics[clip=true,width=0.54\textwidth]{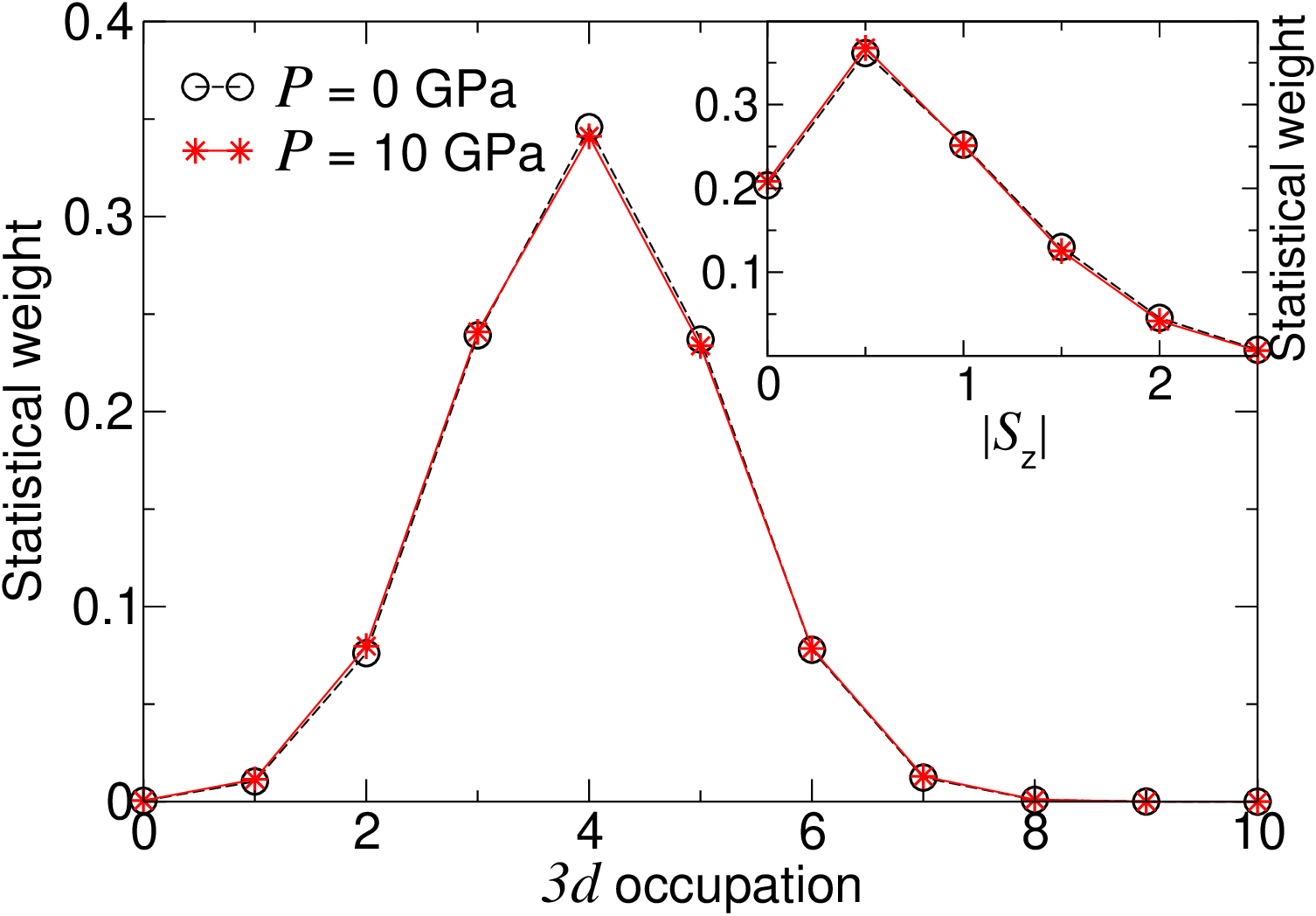}
\caption{\label{fig:sweigths}
Statistical weights of atomic configurations obtained within quantum Monte Carlo at pressures ${p=0, 10}$~GPa and temperature ${T = 193}$~K. Inset: statistical weights of absolute value of z-projection of spin $S_z$.}
\end{figure}

To summarize our study of local electronic and magnetic properties, in figure~\ref{fig:sweigths} we present the statistical weights of various electronic and magnetic configurations. Although the maximum weight is achieved for the electronic configuration with occupation $n=4$ (in agreement with obtained average occupation), this maximum is rather broad, such that the actual occupation of various configurations varies in the range $2\leq n\leq6$. Likewise, the spin projection has maximal weight at $|S_z|=1/2$, but varies in the range $|S_z|\leq 2$. This confirms once more the characterization of chromium as a weak magnet without formed local moments.

\subsection{Non-local magnetic properties}

To calculate the N\'eel temperature, we approximate the magnetic wave vector by ${\mathbf{Q}_{\rm H}}\,{=}\,(0, 0, 2\pi/a)$
and construct a supercell for modeling an antiferrromagnetic order corresponding to this wave-vector. 
In particular, our supercell contains two nearest-neighbor atoms at $(0,0,0)$ and $(a/2,a/2,a/2)$ in Cartesian coordinates.
The corresponding lattice vectors are ${\{a,0,0\}}$, ${\{0,a,0\}}$ and ${\{0,0,a\}}$. 
The constructed supercell allows us to consider two magnetic sublattices and calculate the 
response to a non-uniform magnetic field, which 
site dependence ${\mathbf{H}_{\mathbf{R}_j} = \mathbf{H}_0\, \textrm{cos}(\mathbf{Q}_{\rm H} \mathbf{R}_j)}$ corresponds to the antiferromagnetic order.
In other words, we compute the sublattice magnetic susceptibility $\chi_{\rm sub}$ as a response to a small staggered magnetic field $\mathbf{H}_0$ introduced in the DMFT part.
In the calculations
we have used the magnetic field corresponding to splitting of the single-electron energies up to 5~meV, which was checked to provide a linear response. The magnetic transition temperatures are obtained from the  onset of spontaneous sublattice magnetization.

\begin{figure}[h]
\centering
\includegraphics[clip=true,width=0.57\textwidth]{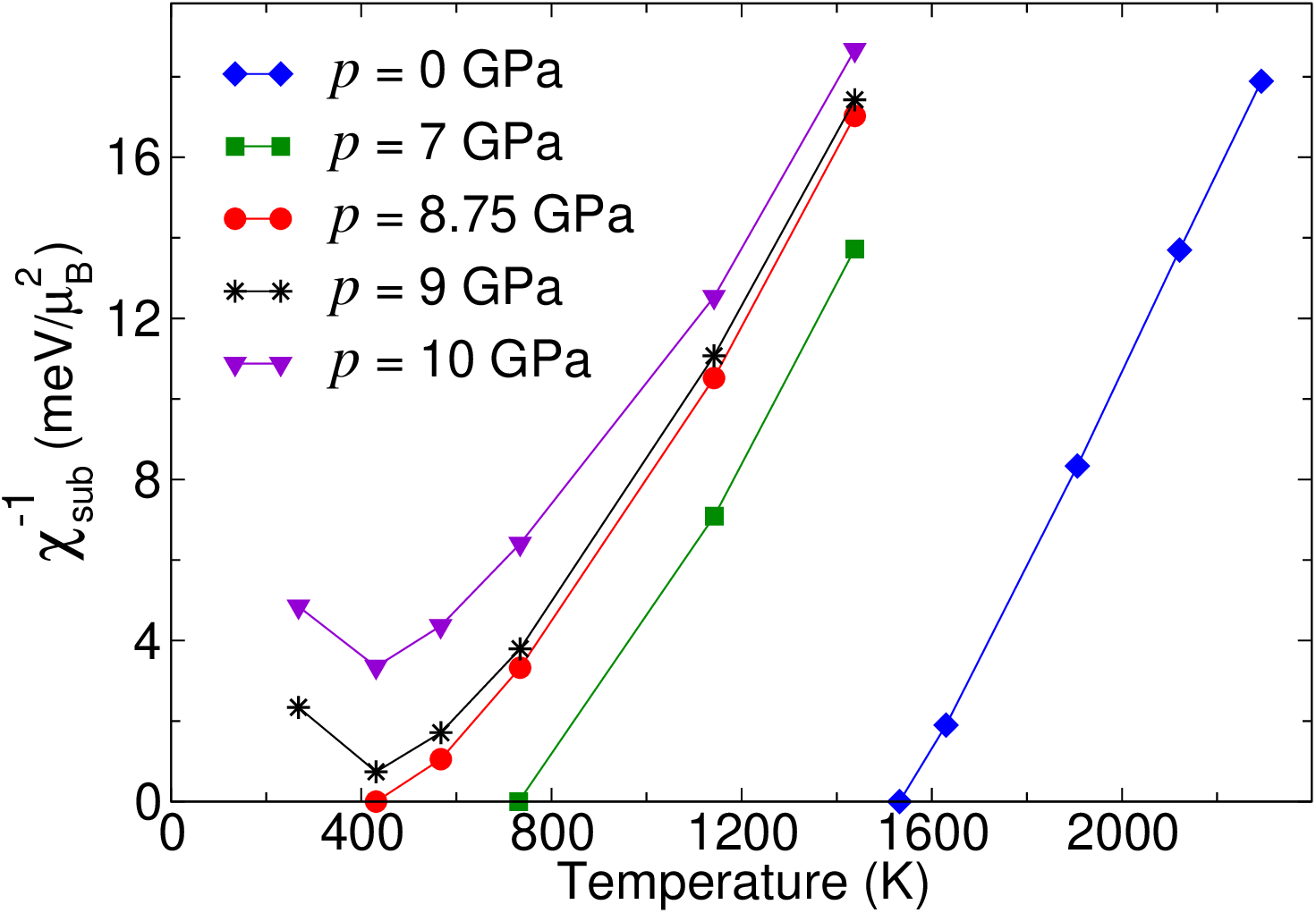}
\caption{ \label{susc}
Inverse sublattice susceptibility as a function of temperature obtained by the DFT+DMFT method at various pressures. The $\chi^{-1}_{\rm sub}=0$ points are obtained from the onset of spontaneous sublattice magnetization.}
\end{figure}

\begin{figure}[b]
\centering
\includegraphics[clip=true,width=0.57\textwidth]{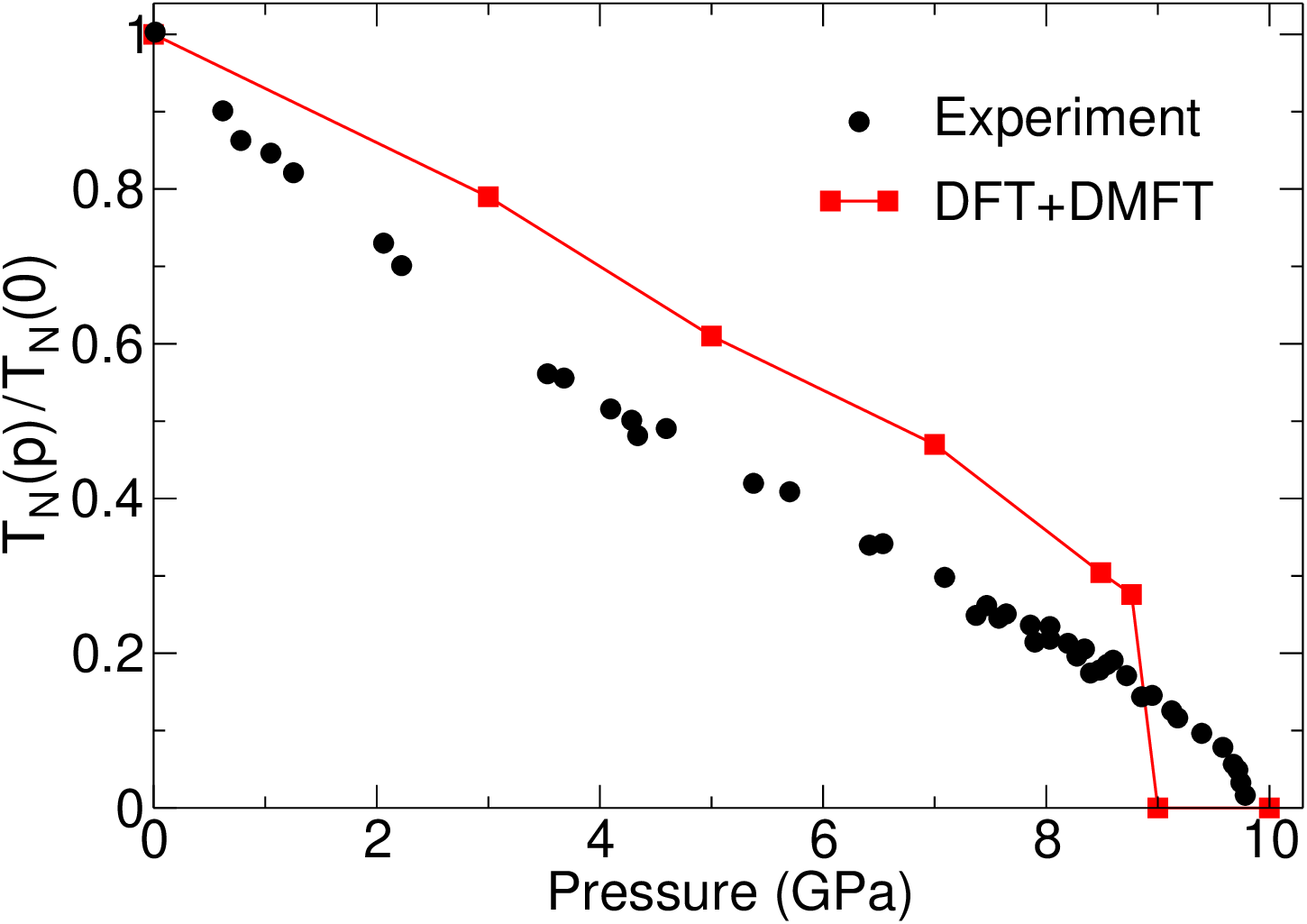}
\caption{ \label{fig:loc_and_uniform_susc}
Relative value of N\'eel temperature as a function of pressure obtained by the DFT+DMFT method in comparison with experimental data~\cite{Pressure2}.
}
\label{FigTN}
\end{figure}

The obtained temperature dependence of inverse sublattice susceptibility and corresponding magnetic transition temperatures are  shown in figure~\ref{susc}.  The obtained N\'eel temperature at ambient pressure (1500~K) is strongly overestimated in comparison with the experimental value $T_{\rm N}\approx 300$~K. This is in a sharp contrast with strong magnets, such as iron \cite{
Fe,alpha_iron2010,OurAlphaIgoshevKatanin,OurAlphaBelozerovKatanin}, where overestimation of Curie temperature is only moderate (1.5 to 2 times) and partly originates from the Ising symmetry of Hund exchange \cite{Hausoel,BelozerovSU2}. The remaining part of overestimation is due to the mean-field nature of DMFT approach. For systems with local magnetic moments, in particular strong magnets, such as iron, this implies mean-field approximation to the effective Heisenberg model (cf. Refs. \cite{OurAlphaIgoshevKatanin,OurAlphaBelozerovKatanin}), which is known to overestimate the Curie temperature by a factor 1.3-1.5 depending on the coordination number. At the same time, in the regime of weak electronic correlations, as obtained in section~\ref{Sect31}, DMFT performs mean-field-like treatment of itinerant magnetic degrees of freedom. In view of the similarity of diagrammatic structure of DMFT magnetic susceptibility to that in RPA approach\cite{OurRev}, the above mentioned mean-field-like treatment is expected to strongly overestimate magnetic transition temperatures similarly to Stoner (Overhauser) theory 
 \cite{Moriya}. 
Yet, some effects of electronic correlations are accounted by the DMFT approach, as can be seen, e.g., from the close to linear temperature dependence of inverse susceptibility far from quantum phase transition. The obtained pressure dependence of reduced N\'eel temperature $T_{\rm N}(p)/T_{\rm N}(0)$ is shown and compared to the experimental data of Ref. \cite{Pressure2} in figure~\ref{FigTN}. Similarly to the experimental data, the obtained dependence $T_{\rm N}(p)$ is almost linear in a broad pressure range, and then sharply drops to zero at a critical pressure ${p_c=9}$~GPa, which is close to the experimental value of $9.5$~GPa. The obtained sharp drop of $T_{\rm N}(p)$ indicates a first-order quantum phase transition in the considered supercell DMFT approach.

\begin{figure}[h!]
\centering
\includegraphics[clip=true,width=0.54\textwidth]{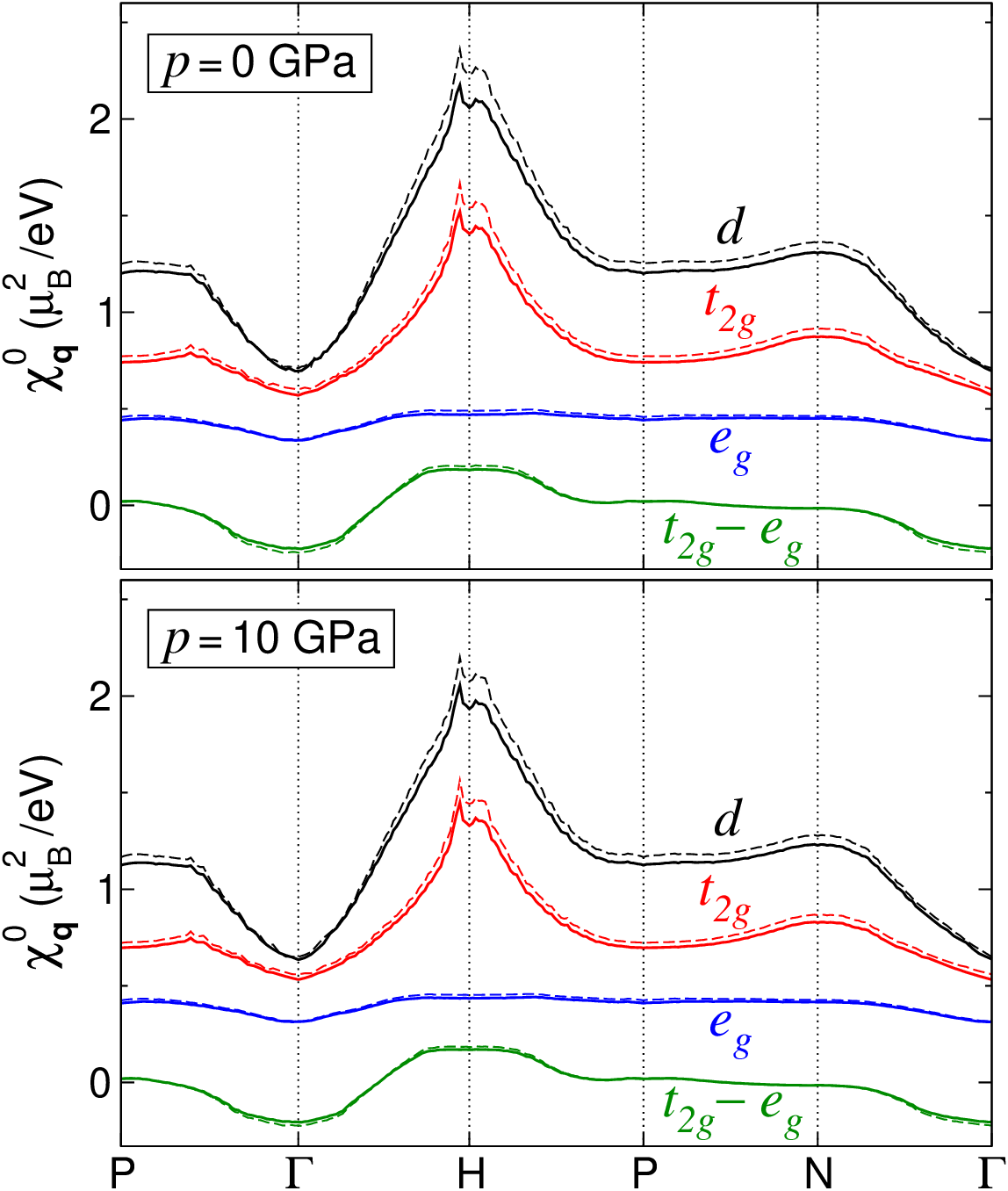}
\caption{\label{fig:susc_irr}
Momentum dependence of the particle-hole bubble
(Eq.~\ref{Eq:bubble})
calculated within DFT (dashed lines) and DFT+DMFT (solid lines)  under ambient pressure (top panel) and ${p = 10}$~GPa (bottom panel) at temperature ${T=193}$~K. The contribution from all $d$ orbitals (black lines) and partial $t_{2g}$ (red lines), $e_g$ (blue lines), and $t_{2g}$-$e_g$ contributions (green lines) are present.}
\end{figure}

\begin{figure}[h!]
\includegraphics[clip=true,width=0.57\textwidth]{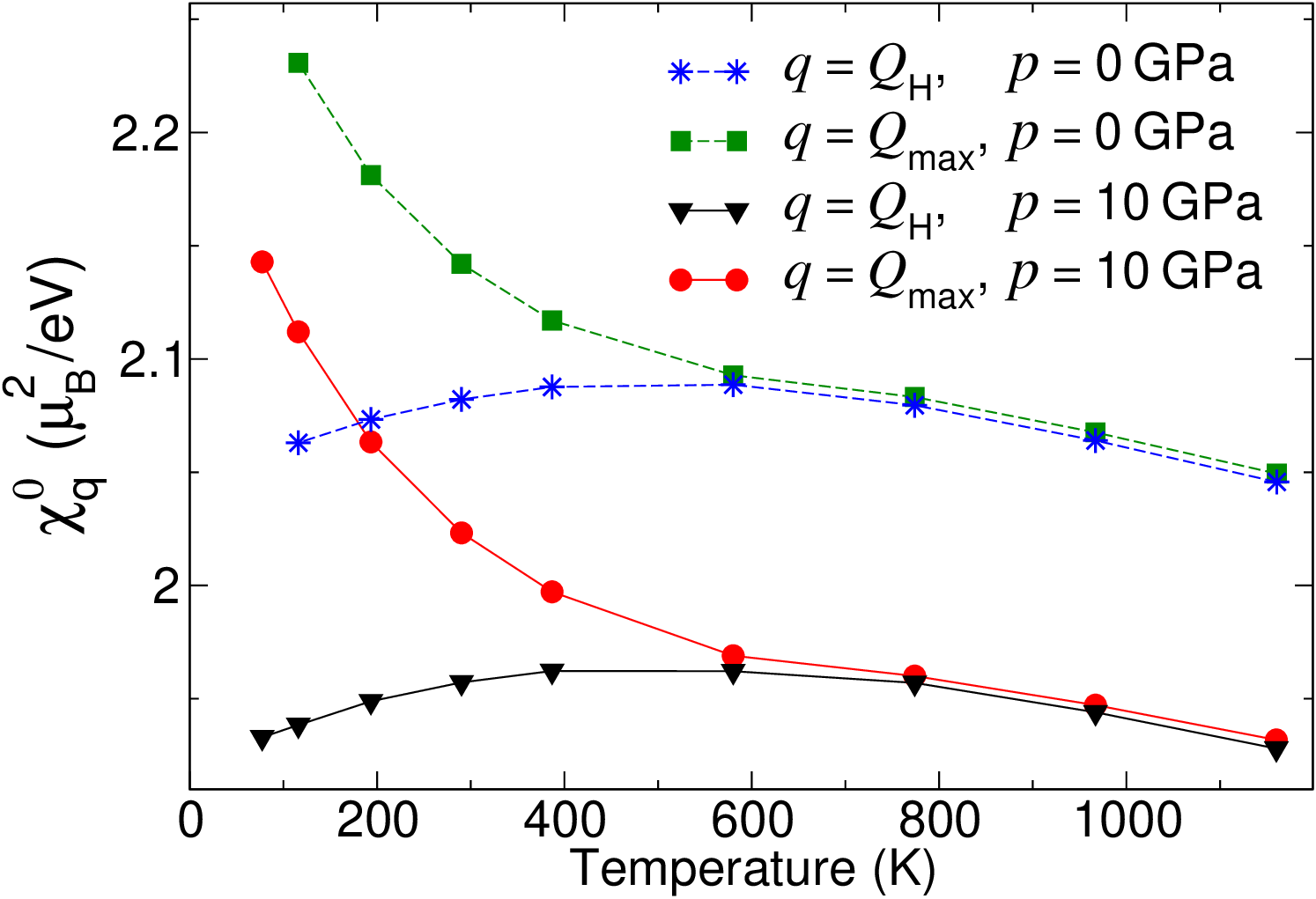}
\centering
\caption{\label{fig:susc_QH}
Temperature dependence of the particle-hole bubble (Eq.~\ref{Eq:bubble})
at the staggered wave vector $\bf{Q}_{\rm H}$, calculated within DFT+DMFT at pressures ${p = 0, 10}$~GPa, compared to that at the wave vector of the maximum of the bubble $\bf{Q}_{\rm max}$.}
\end{figure}



To understand the role of magnetic correlations with various wave vectors, we
%
study the momentum-dependence of static (zero frequency) magnetic susceptibility. In particular, we compute the lowest-order (with respect to the vertex corrections) non-uniform magnetic susceptibility, which corresponds to the particle-hole bubble diagram and can be written as
\begin{equation}
\chi_{\bf q}^{0} = -(2\mu_B^2/\beta) \sum_{{\bf k},\nu_n} \textrm{Tr}\left[G_{\bf k} (i\nu_n) G_{\bf k+q}(i\nu_n)\right],
\label{Eq:bubble}
\end{equation}
where $\mu_B$ is the Bohr magneton, $\nu_n=(2n+1)\pi T$ are the fermionic Matsubara frequencies, the trace is taken over the orbital indices of Wannier functions of $d$ symmetry (for more details see Refs. \cite{dmft1,OurRev}).
In figure~\ref{fig:susc_irr} we show the momentum dependence of $\chi_{\bf q}^{0}$ calculated using non-interacting (DFT) and interacting (DFT+DMFT) Green's functions. One can see that sharp peak near the wave vector ${\mathbf Q}_{\rm H}$, obtained previously in DFT at ambient pressure \cite{Susc0,Susc1,OurCr}, is preserved in DMFT, which accounts for the interaction effects, similarly to previous study of the single-band Hubbard model \cite{OurKohn}. As it was discussed in Refs. \cite{Roth,Rice,OurKohn,OurCr}, these sharp peaks originate from Kohn points (or lines) of the Fermi surfaces. These are the points (or lines), connected by the wave vector ${\mathbf Q}_{\rm max}$, corresponding to the peak position, and having opposite Fermi velocities in two perpendicular directions. Since the sharp maximum of the susceptibility is preserved with changing pressure, these points (or lines), corresponding to ``local nesting" regions, are not destroyed with applying pressure (see also Ref. \cite{OurCr1}).

The obtained shape of the momentum dependence of the bubble allows one to explain the first order phase transition, obtained in the considered supercell DMFT approach with applying pressure. For that, we note that the first order transition occurs since the minimum of the dependence $\chi^{-1}_{\rm sub}(T)$ is not shifted to ${T=0}$~K with applied pressure (see figure~\ref{susc}). At the same time the staggered susceptibility decreases, such that at the critical pressure $p_c$ the solution of the equation $\chi^{-1}_{\rm sub}(T)=0$ disappears. 
To understand the origin of this behavior, in figure \ref{fig:susc_QH} we plot the temperature dependence of the bubble at the pressures $p=0$ and $10$~GPa and wave vectors ${\bf q}={\bf Q}_{\rm H}$ and ${\bf Q}_{\rm max}$.
One can see that the temperature dependence of the bubble at the wave vector ${\bf Q}_{\rm H}$ is also non-monotonic, similarly to the staggered susceptibility of figure \ref{susc}. This behaviour is natural, since the non-local susceptibility is related to the frequency-resolved bubble via the Bethe-Salpeter equation \cite{dmft1}, containing local vertices. At the same time, the bubble at the wave vector ${\bf Q}_{\rm max}$ shows almost linear monotonic behavior at small temperatures, which is typical for systems with Kohn points (lines) at the Fermi surface \cite{Rice,OurCr,OurKohn}, and therefore it is substantially enhanced at low temperatures. This allows us to conclude that in the corresponding temperature range $T\lesssim T_N(0)/3$ the incommensurability of magnetic order, not accounted in the considered supercell approach, becomes essential. Therefore, we expect the replacement of the obtained first-order transition by a second-order transition (happening at higher pressures) from incommensurate to paramagnetic phase, when incommensurate correlations are taken into account. A similar result was found in the previous static mean-field and slave boson studies of two- \cite{Q2D} and three-dimensional \cite{Q3D} systems. 
%

\section{Conclusions}

In summary, we have studied the electronic and magnetic properties of chromium within DFT+DMFT approach. In electronic and local magnetic properties we find a quite weak effect of electronic correlations: the self-energies have Fermi-liquid like form, local static magnetic susceptibilities are only weakly temperature dependent, and the real part of local dynamic susceptibility has a broad plateau as a function of real frequency. 
This is in contrast to such $3d$ metals as vanadium, iron, cobalt, nickel, which are further from the half-filling of $d$-states than chromium, but nevertheless show much stronger many-body effects. The reason for the unusually weak correlations in chromium may be in its electronic structure. Namely, the DOS at the Fermi level $N(E_{\rm F})$ has a dip, that reduces the number of particle-hole excitations. Moreover, peaks of DOS, which may be a source of stronger correlations,
are quite distant from the Fermi level in chromium.
A somewhat 
similar to chromium shape of the DOS near the Fermi level is observed in
the $\varepsilon$-phase of iron.
At the same time, $\varepsilon$-iron, in contrast to chromium, has short-lived local magnetic moments~\cite{epsilonFe}, which likely occur because of a larger $N(E_{\rm F})$ and a closer position of the peak of DOS to the Fermi level. The local moments in $\varepsilon$-iron are however still only weakly formed and they are difficult to detect experimentally.

Our study of staggered susceptibility of chromium shows its non-monotonic temperature dependence, which is also present in the particle-hole bubble. The calculated N\'eel temperature decreases approximately linearly with applied pressure, in agreement with the experimental data. We find a first-order magnetic transition at applied pressure ${p\sim 9}$~GPa. As we argue, however, this first-order transition occurs because of the above mentioned non-monotonic dependence of staggered susceptibility and
neglect of incommensurate magnetic correlations, which can not be treated easily in considered supercell approach. We expect that account of these correlations will yield the second-order phase transition. 

In the light of these results, further studies of the non-local effects in chromium are required. This can be performed, e.g. via the calculation of non-uniform susceptibility in non-local extensions of DMFT \cite{OurRev}. In view of the required description of low-temperature behavior of susceptibilities near quantum phase transition, these calculations should be supplemented by efficient impurity solvers, which are able to calculate vertex functions of multi-orbital models in the low temperature regime. The suppression of the N\'eel temperature with respect to the results of DMFT can be also further studied in the recently proposed approach \cite{Licht}. 

\ack
The authors are grateful to D Volkova for the help with Wannier functions and also to D~Gazizova and S~Streltsov for useful discussions. The DFT+DMFT calculations were supported by Russian Science Foundation (project 19-12-00012).
The calculations of the particle-hole bubble were supported by the Ministry of Science and Higher Education of the Russian Federation (theme “Electron” No. AAAA-A18-118020190098-5).


\begin{thebibliography}{99}
\bibitem{footnote1}
The spin-spin correlation function for $\alpha$-iron was calculated by the DFT+DMFT method at inverse temperature ${\beta=7}$~eV$^{-1}$ with interaction parameters ${U=4}$~eV and ${J_{\rm H}=0.9}$~eV. Details of calculations are presented in reference~\cite{OurAlphaBelozerovKatanin}.
\end{thebibliography}
\end{document}